# Nanoscale single- and multi-photon mapping of optical fields

David Bitauld[1*†], Francesco Marsili[1,2‡], Alessandro Gaggero[3], Francesco Mattioli[3], Roberto Leoni[3], Saeedeh Jahanmiri Nejad[1], Francis Lévy[4], and Andrea Fiore[1]

[1]COBRA Research Institute, Eindhoven University of Technology (TU/e), PO Box 513, 5600 MB Eindhoven, The Netherlands

[2]Institute of Quantum Electronics and Photonics (IPEQ),Ecole Polytechnique Fédérale de Lausanne, Station 3, CH-1015 Lausanne, Switzerland

[3]Istituto di Fotonica e Nanotecnologie (IFN), CNR, via Cineto Romano 42, 00156 Roma, Italy

[4]Institute of Condensed Matter Physics (IPMC), Ecole Polytechnique Fédérale de Lausanne, Station 3, CH-1015 Lausanne, Switzerland

[*]e-mail: david.bitauld@tyndall.ie

[†]Current address: Photonics Theory Group, Tyndall National Institute, Lee Maltings, Prospect Row, Cork, Ireland

[‡]Current address: Department of Electrical Engineering and Computer Science, MIT, Cambridge, MA 02139, USA

The study of optical phenomena on the subwavelength scale is becoming increasingly important in photonics, particularly in the fields of nanoemitters, photonic crystals and plasmonics. Subwavelength field patterns are evanescent and must thus be investigated with near-field techniques. The light powers emitted by nanoscale sources are extremely low, undermining the traditional approach of scattering a near field to a large (thus noisy) far-field detector. Nanoscale detectors, providing direct sensing in the near-field with small noise due to a small active area, are needed in high-sensitivity, high-resolution near-field imaging and in quantum nanophotonic circuits. Here we report the first nanoscale (≈50×50 nm$^2$) detector displaying single-photon sensitivity and a nanosecond response. These nanodetectors can also be operated in multi-photon mode, where the detection threshold can be set at N=1, 2, 3 or 4 photons, thus allowing the mapping of photon number statistics on the nanoscale.

In order to overcome the sensitivity-resolution trade-off typical of near-field microscopy, the integration of nanoscale detectors on scanning tips was proposed[1,2] and some approaches to the fabrication of nanodetectors have been demonstrated[3,4,5,6,7] but their sensitivity has remained far from the single-photon regime. Our single-photon nanoprobe is based on the principle of photon-induced hot-spot formation in superconducting nanowires, which also underpins superconducting single-photon detectors (SSPD)[8] and photon-number-resolving detectors[9]. Photons absorbed by a thin superconducting film produce a local resistive region ("hot-spot") of few tens of nm in diameter. We pattern a single nanoscale constriction in the wire (Fig. 1), and bias it close to the constriction's critical current. When a photon is absorbed in the constriction, the superconducting current is expelled from the hot-spot and redistributes in the still superconducting sidewalks, where the current density increases. As the critical current density is exceeded, a resistive barrier is formed across the entire constriction, producing a measurable voltage pulse. The detectors have a spatial resolution close to the constriction size, since hot-spots created outside the constrictions are not detected due to the lower current density. This single-photon regime is obtained when the bias current is close to the critical current. For lower bias currents, a single hot-spot is not sufficient to trigger detection, and the detector responds only to two or more photons, which, as shown below, can be used for multiphoton detection.

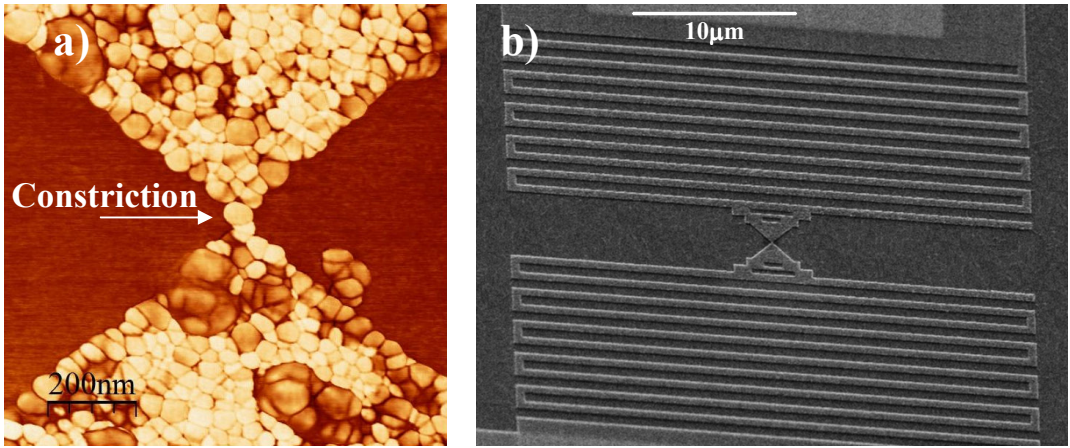

Figure 1: Nano-detector design. a) Atomic-force microscopy (AFM) image of the nanoscale detector. The narrowest part of the device constituting the active region is 50 nm ± 4 nm wide. The structure was imaged without stripping the electron resist, which produces the observed roughness. b) SEM image showing the whole device including the meander inductors on each side of the detector.

The nanodetectors were fabricated from 4 nm-thick NbN films (see Methods). As the kinetic inductance of the nanoscale constriction is very small, we added additional series inductors (two 285μm-long meanders in Fig. 1b) to avoid latching in a permanent resistive state (see Yang et al.[10] and Methods section). We first tested 50 nm detectors in the single-photon detection regime. In order to demonstrate that each detection pulse arises from the absorption of a single photon, we measured the count rate as a function of the incident power for a constant bias current of $I_b=5\mu A$ (Fig. 2a), close to the critical current $I_c=7\mu A$. The dependence of the counts as a function of power is linear indicating that a single photon process is involved. For the 50 nm-wide device studied here this linear dependence was observed for any bias current in the measurable 3.5-7 µA range, which indicates that the diameter of the hot-spot is close to the constriction size. The amplified photodetection pulse is displayed in the inset of Fig. 2a, showing a 1/e decay time of 2.5 ns, close to the calculated value L/R=2 ns (see Methods). While we did not directly measure the jitter, we expect values <100 ps, similar to meander SSPDs[11].

The count rate under illumination ($\lambda=1.3\mu m$, 26MHz repetition rate, 850 photons/pulse in total, 0.2 photons/pulse on the expected active area) as well as the dark count rate as a function of the bias current are displayed in Fig.2 b). For this measurement as well as every measurement presented in this paper the temperature of the sample was 6K. Due to the stronger current dependence of the dark count rate, a signal-to-noise ratio of five orders of magnitude is obtained at 4.3 µA, even at the relatively high measurement temperature of 6K (at 4.2 K a much lower dark count rate was measured). The corresponding noise equivalent power is NEP=$10^{-17}$ W.Hz$^{-1/2}$ (several orders of magnitude better than previously reported values for nanodetectors: $1.5\times10^{-13}$ W.Hz$^{-1/2}$ in ref. 3 and $7\times10^{-15}$ W.Hz$^{-1/2}$ in ref. 6). The photocounts vs bias current curve is relatively flat in the range $4.3\mu A<I_b<7\mu A$ as compared to meander SSPDs fabricated with the same technology (the sharp decrease of the count rate below 4.3 µA in Fig. 2b is in fact due to the pulses becoming too small and thus not being detected by the counter). This holds true even for illumination up to 2.05 µm (the largest wavelength that could be tested), showing the potential for efficient near-field detectors even in the mid-infrared.

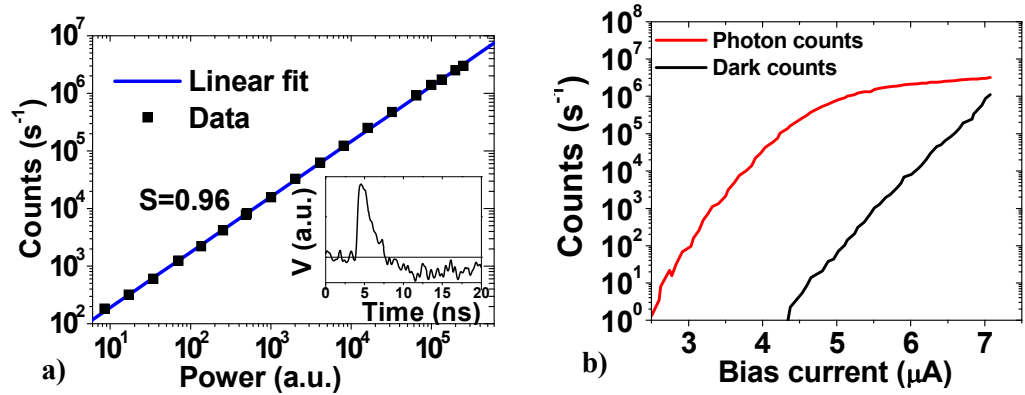

Figure 2: Photoresponse counts as a function of power and current. a) The number of counts as a function of the light power illuminating the device ($\lambda=1300$ nm) displays a linear response on 5 orders of magnitude. Inset: Photoresponse transient. b) Counts as a function of the bias current of the device under illumination (photon counts) and in the dark (dark counts). Light was polarised parallel to the direction of current flow.

By calculating the number of photons incident on the geometrical area of the detector (~50×50nm$^2$), we derive from Fig. 2b a nominal value of the quantum efficiency, with a maximum of 58% at $I_b$=7μA even higher than the estimated ~28% absorption of the 4 nm-thick NbN film[12]. This surprising value can be due to an underestimation of the effective active area. Indeed, the areas surrounding the constriction also become sensitive to light at large bias current, thereby increasing the active device area. Based on the ratio between measured efficiency and nominal absorption, we estimate an effective area ~70×70nm$^2$ at $I_b$=7μA. The best spatial resolution will be then obtained at slightly lower bias currents or by using tapers with larger angles.

We then tested the spatial resolution, by scanning the illumination spot over the detector in two dimensions and measuring the photocounts as a function of position (Fig. 3). This spot is produced by focusing a parallel beam (λ=780nm) with a reflective objective. It is composed of a 1μm central spot (FWHM) and four 500nm sidelobes. This unusual diffraction spot is due to the used reflective objective, which presents some obscuration areas. We can see that the detector can image features as small as the 500nm side lobes of the spot – this is an upper limit to the spatial resolution that we expect to be in the ≈100 nm range. In order to directly show the detector sensitivity, the experiment was repeated attenuating the laser to an average of 0.5 photons/pulse (inset in Fig. 3). A clear diffraction pattern with large signal-to-noise ratio is observed, proving the detector's unique ability to directly measure the diffraction of single photons with submicrometer resolution.

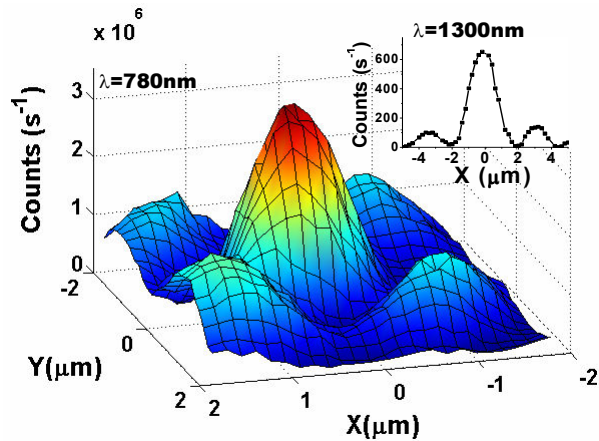

Figure 3: 2D scan of the photocounts as a function of the light spot position (λ=780 nm, number of photons per pulse >>1). Inset: 1D scan (λ=1300 nm and 0.5 photons per pulse in average).

When larger constrictions (e.g. 150 nm) are used, the absorption of a single photon is not always sufficient to make the entire constriction resistive, as the hot-spot size is smaller than the constriction. By choosing the bias current, we set whether one, two, or more photons are necessary for detection. Figure 4 displays the count rate R of a 150 nm-wide detector ($I_c$=21 μA) as a function of the average number μ of photons per pulse in the illuminating spot for different currents, chosen to obtain a $R \propto \mu^s$ dependence (s=1-4). The fitted slopes have almost integer values showing that the detector fires for N=1-4 photons (the detector can also fire for >N photons absorbed in the active area, but such events are rare in the used range of illumination power). We note that, while multiphoton detection is also observed in meander SSPDs[13], it is usually extremely inefficient, since the probability that the photons incident on the 10x10 $\mu m^2$ meander area are absorbed in close proximity is very small. In contrast, in the nanodetector multiple photons incident on the 150×150 $nm^2$ active area have a high probability of producing a multiphoton detection (see Methods).

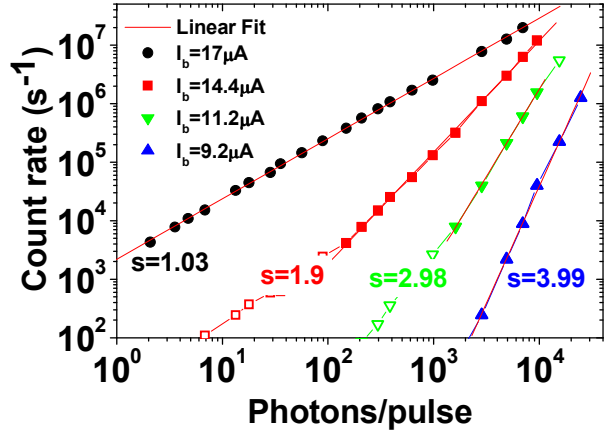

Figure 4: Count rate as a function of the average number of photons per pulse for four different bias currents in a 150 nm-wide detector. The pulse rate was 80MHz. The measured points represented by full symbols were fitted with straight lines in log-log scale. Their respective slopes are displayed on the graph. At low average photon numbers the experimental points (empty symbols) deviate from the ideal slope since the probability of N-photon pulses becomes negligible.

A 1D scan of the light spot was performed for the four bias currents (Fig. 5) ($\lambda$=780nm, $10^4$ photons per pulse). In the multiphoton regime, the width of the spot is reduced as $1/\sqrt{N}$ thus increasing the contrast of the image as compared to single-photon or classical intensity imaging[14] (the 1/N dependence observed in other quantum imaging experiments requires cancelling of lower-order interference effects, e.g. through the use of entangled-light illumination[15]). This nonlinear imaging using a quantum detector is demonstrated here for the first time on the subwavelength scale.

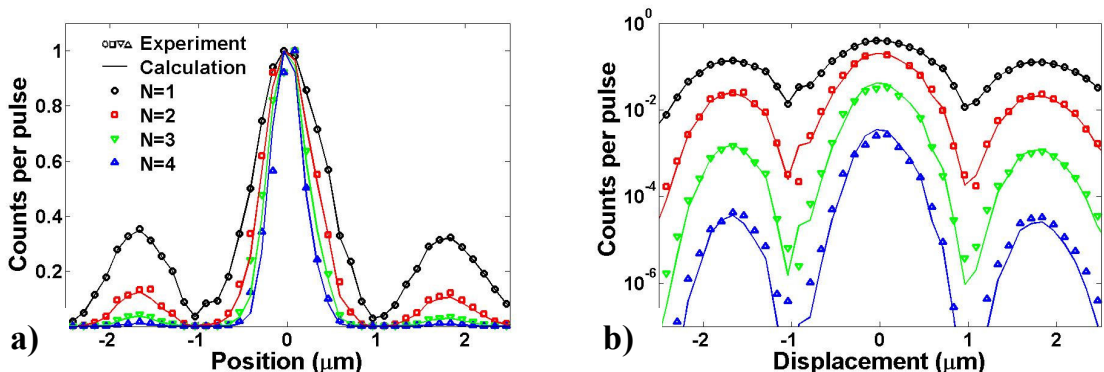

Figure 5: Measured counts (symbols) as a function of the light spot position for the bias currents corresponding to N-photon regimes (N=1-4), and related fits (lines). (a) Normalised and in linear scale, and (b) as measured, in log scale.

In order to verify that the profiles displayed by the four curves are consistent with multiphoton detection, we used a statistical model to calculate the detection probability for the different N-photon regimes (see Methods). The average number of photons absorbed by the active area as a function of position was derived from the single-photon counts, and then used to calculate the expected detection rate in the three other regimes for each position (continuous lines for N=2-4 in Fig. 5a and b). We obtain an excellent agreement with the experimental scans over several orders of magnitude, confirming the accuracy of this simple model.

In conclusion, we have demonstrated the first nanoscale single-/multi-photon detector and used it as a probe to map an optical field with submicrometer features. We were able to measure the diffraction pattern of single photons with subwavelength spatial resolution. Multi-photon discrimination was also demonstrated, paving the way to a resolution beyond the diffraction limit using non classical light, i.e. to performing quantum imaging experiments on a subwavelength scale. This detector could also be directly applied to near-field microscopy. Indeed, by fabricating a nanodetector on top of a sharp tip (e.g. on silicon, on which high-quality NbN films can be grown[16], and using fabrication techniques similar to those developed for scanning thermal microscopy[20]) and scanning it in the near field of an emitting sample, a near-field map may be obtained with ≈50 nm spatial resolution, single- and multi-photon sensitivity, and <100 ps temporal resolution, opening novel opportunities in the field of low-temperature near-field microscopy of semiconductor photonic and electronic nanodevices. Finally, our nanoscale detectors could be integrated with nanoscale sources and optical circuits, to provide the required ultrasensitive detection functionality in integrated nanowire photonics[6] and (single-)plasmonics[18,19].

METHODS

FABRICATION

The fabrication process has been described in Marsili et al.[17]. In short, 4 nm thick films of NbN were grown on MgO substrates by reactive magnetron sputtering in an argon–nitrogen gas mixture. The substrate temperature was 400°C. The films exhibited a superconducting transition temperature of 10.5 K and a superconducting transition width of 0.3 K. The device is composed of a constricted region (50 or 150 nm-wide) yielding the active area and two inductors. Indeed, as the kinetic inductance is very small (≈90 pH), the current returns to the wire within few ps (L/R=1.8 ps), when the hot-spot is not yet completely healed. In this situation, ohmic heating can keep the constriction in the resistive state indefinitely ("latching", ref. 10). In order to avoid latching, we placed an additional inductance, in the form of two 285 μm-long, 500 nm-wide meanders, in series with the constriction (Fig. 1b), resulting in a total inductance $L_{kin}$=100 nH and an estimated recovery time L/R=2 ns (R=50 Ω, the impedance of the transmission line). The critical current of this inductor is ten times larger than that of the active area, so that at the bias currents used in the experiment photons absorbed in the inductor are not detected. Nanolithography was carried out using a field-emission gun electron-beam lithography system (acceleration voltage 100 kV). In a first step, pads and alignment markers (60 nm thick gold on 10 nm thick titanium) were fabricated by lift-off using a polymethylmethacrylate (PMMA) stencil mask. In a second step, a hydrogen silsesquioxane (HSQ) mask was defined, reproducing the pattern of the device. All the unwanted material, i.e. the material not covered by the HSQ mask and the Ti/Au film, was removed by using fluorine-based reactive-ion etching. An AFM micrograph of the active area is shown in Fig. 1a. The irregularities of the surface are due to the HSQ layer remaining on top of the NBN film. Fig. 1b is an SEM micrograph of the complete device displaying also the inductors.

CHARACTERIZATION

The characterization of the detector was carried out in a cryogenic probe station with an optical window. The temperature of the sample's surface in this cryostat was about 6 K. In Fig. 2, the device was illuminated through a long working distance (34 mm) objective with numerical aperture NA=0.3 producing a Gaussian spot with a waist w=2.6±0.1 μm at 1300 nm. In Fig. 3-5, we used a high numerical aperture (NA=0.4) long working distance (24mm) reflective objective, producing a ≈1 μm-diameter spot at 780nm. The light was produced by pulsed lasers (λ=780 or 1300 nm) and guided to the optical setup with single-mode fibers. The optical setup was mounted on an XYZ translation stage with 50 nm steps allowing us to optimize the focus in Z and perform scans in X and Y. The electrical contact was performed with high frequency microprobes. The bias current was provided by a low-noise voltage source through a series resistance of 10Ω and the DC port of a bias-tee. The electrical pulses produced by the detector were retrieved through the RF port of the bias-tee, amplified with wide-bandwidth amplifiers and directed to a counter.

MODEL

We consider that the photons incident on the nominal detector area have a probability η to be absorbed in a region where they can contribute to the photodetection process (η may depend on the current and on the geometry of the constriction) . If the number of absorbed photons is superior or equal to N, a N-photon detection is triggered. The probability of having k absorbed photons out of m incident photons is given by the binomial law:

$$E(k/m) = \binom{m}{k} \eta^k (1-\eta)^{m-k}$$

The probability of having N or more absorbed photons (i.e. $k \geq N$) is equal to 1-Prob($k < N$):

$$P_N(m) = 1 - \sum_{k=0}^{N-1} E(k/m) .$$

In our case the photon number per pulse is determined by a Poissonian law with a mean photon number µ. The detection probability is then:

$$D_N(\mu) = \sum_{m=N}^{\infty} P_N(m) \times \frac{\mu^m e^{-\mu}}{m!} \quad (1).$$

In order to avoid the numerical difficulty to calculate factorials of large numbers we used Gosper's approximation $m! \approx \sqrt{\left(2m + \frac{1}{3}\right)\pi} m^m e^{-m}$.

The average number of photons µ(x) incident on the device active area (150×150 nm$^2$) is a function of the spot position x. When the light spot is centred (x=0), $\tilde{\mu}$ can be estimated from the measured image of the laser spot (Fig. 3) which gives ~64 incident photons on average out of 10$^4$ photons in a pulse for a surface of 150×150 nm$^2$. From the number of counts at x=0 in the single-photon regime (N=1), we derive η(I$_1$)=0.8%. The average number of absorbed photons µ(x) is then calculated as a function of position from the N=1 scan, using expression (1). This µ(x) profile is then used for plotting the expected detection probability $D_N(\mu(x))$ for N=2,3 and 4 (continuous lines in Fig. 5a and 5b), using η(I$_N$) as a single fitting parameter for each curve (η(I$_2$)=1.3%, η(I$_3$)=1.2% and η(I$_4$)= 0.97%). The fitted efficiencies are nearly equal, which confirms our assumption that the number of photons absorbed in the active area is the key parameter driving the N-photon detection.

ACKNOWLEDGEMENTS

The authors thank H. Jotterand for technical support. This work was supported by EU FP6 STREP 'SINPHONIA' (contract no. NMP4-CT-2005-16433), the Dutch NRC Photonics program, the Swiss National Science Foundation through NCCR Quantum Photonics program, and the EU FP6 IP 'QAP' (contract no. 15848). A.G. gratefully acknowledges a PhD fellowship at University of Roma TRE.